\documentclass[reprint,aps,prstper]{revtex4-2}

\usepackage{graphicx}
\usepackage{hyperref}
\usepackage{footnote}
\usepackage{amsmath,amssymb}
\usepackage{multirow}
\usepackage{tabularx}
\usepackage{array}
\usepackage[utf8]{inputenc}
\usepackage{enumitem}
\usepackage{subfig} 

\begin{document} 

\title{Lessons for instructor-student interactions in physics from the world of improvisational theater}

\author{Colin G. West$^1$}
\email[]{colin.west@colorado.edu}
\affiliation{$^{1}$Department of Physics, University of Colorado, Boulder, Colorado 80309, USA}

\date{\today}

\begin{abstract}
A considerable share of the literature on physics education and on education more broadly focuses on the principles which should guide the design of courses and of classroom activities. In this short article I wish to place more attention on the unplanned aspects of teaching: specifically, the spontaneous interactions that occur between instructors and students in settings like office hours, recitations, and when students ask questions during lecture. Because by their nature these interactions require thinking on one's feet, and depend upon the interplay between instructor and student, they share many characteristics with improvisational theater (improv). I document three foundational principles from improv literature (active listening, ``yes-and," and the ``button") and describe how they relate to established principles from physisc education research. I provide examples of how each principle can be used to bring solid pedagogical practices to the unstructured space of instructor-student interactions. 
\end{abstract}

\maketitle


\section{Introduction} 
During my time in graduate school, I was fortunate to spend time working with the Alan Alda Center for Communicating Science at Stony Brook University~\cite{one}. The Center’s most notable feature is its use of techniques from improvisational theater (``improv")~\cite{two}  to develop scientists’ public communication skills—an approach which has been shown to be fruitful in several contexts, including mathematics~\cite{three} and clinical medicine~\cite{four}. The improvisational concepts I learned there and elsewhere have profoundly influenced my approach to interacting with my students. After all, when a student asks a novel question, I do not know where they are coming from nor what preconceived ideas they may bring to the interaction. In this unplanned space our conversation could easily become a meandering mess, but my goal instead is that it should have certain essential elements which make it complete and conceptually satisfying, like a successful improv scene. In this article, I summarize three basic improv concepts and provide examples of their use in the physics classroom: Active listening, the ``Yes-And" principle, and the ``Button." These principles embody widely-studied theories of physics education yet provide succinct, actionable reminders of how to put these theories into practice.

\section{Lesson 1: Active Listening}

The nature of improv requires participants to listen actively and be responsive to the ideas provided by other actors~\cite{five}. This means not merely ``paying attention" but also specifically (``actively") looking for possible points of confusion, and inviting clarification as needed. Imagine the dissatisfying result if two actors begin creating a scene where they discuss their intention to attend a ``rally," but one believes she is a spectator at a motorsports event and the other that he is engaged in political protest.  

This concept carries over virtually unchanged to the teacher-student interaction in a physics classroom. A student discussing a new family of concepts, for instance, cannot be expected to use the canonical terminology correctly every time. But their failure to do so may impede conversations with experts, which in turn makes it harder for them to gain clarity on the terms and concepts, threatening a vicious cycle. It is a task for the instructor, actively seeking out hidden meaning and content, to find places to break through. Once one can identify what a student is expressing, even in vague or nonstandard terms, well-documented educational strategies such as concept substitution~\cite{six} can be employed, with clarity of language leading to clarity of concepts. 

\begin{figure}
    \centering
    \includegraphics[width=0.45\textwidth]{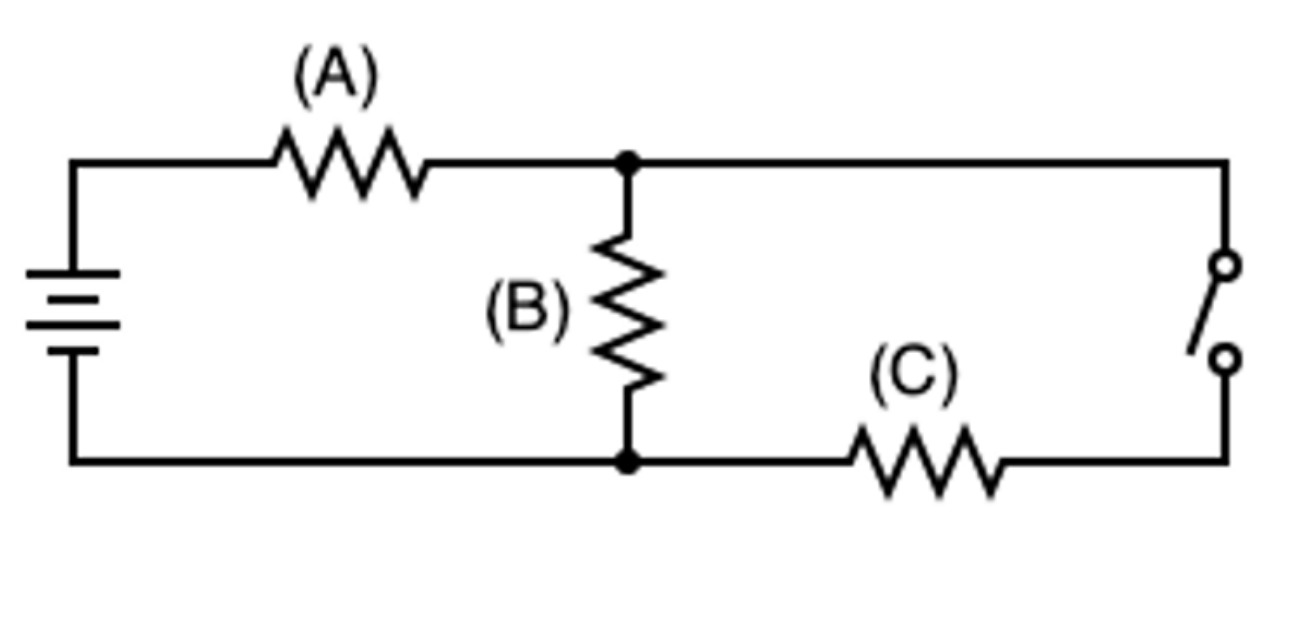}
    \caption{A simple circuit used in a physics problem. Told that all three resistors were identical, Students were asked which was greater: $V_A$ (the potential difference across resistors A), or $V_B$ (the potential difference across resistor B)}
    \label{fig:circuit}
\end{figure}

Consider a conversation I had with a student regarding the circuit in Figure~\ref{fig:circuit}. The student had been asked which was greater: $V_A$ (the potential difference across A), or $V_B$ (the potential difference across resistor B). 

The student told me (paraphrased) that $V_A$  must be greater than $V_B$ because ``to figure out the potential, the key is that some of it goes away by the time we get to resistor B" and that ``it gets used up as you go around the circuit, even if the switch is open."

Instead of clarifying the ambiguous pronoun ``it," I made an initial assumption that the student was talking about current, since students often asserted that current gets ``used up" in resistors. After all, I had several tricks up my sleeve designed to address this common point of confusion.

Unfortunately, my eagerness to jump in had closed my ears to the actual ideas coming from my scene partner. But the key concept was present in the student’s first statement: the student was correct that some of ``the potential" goes away (or ``drops") as one travels across resistor A. The student’s real confusion was about the difference between the terms ``potential" and ``potential difference." Once we had identified the source of the miscommunication, we improvised a very satisfying discussion about the similarities and differences between those concepts. 

\section{Lesson 2: The ``Yes-And" principle}
Perhaps the most celebrated idea in improv is the concept of saying ``Yes-and"~\cite{seven}. The concept, now applied in everything from psychoanalysis~\cite{eight} to the hospitality industry~\cite{nine}, contains two equally important parts: when one’s scene partner introduces a new idea, one must always accept it (``yes!") and must add to it in a way that moves the scene forward (``and"). If an actor says, ``I sure hate a rainy day like this," a ``yes, and" response might be ``Well, that’s Seattle for you. Why did we decide to have our destination wedding here?" By contrast, there is no generative value in replying, ``Oh, that’s just the sprinklers. Now, as I was saying…" 

In improv, the impulse to ``block" another’s contribution arises when an actor cannot tell where the other is going, or if they feel they have a better way to proceed~\cite{two}. In the same way, many instructors instinctively shut down interactions with a student if they are not sure where the student’s question is coming from, or if they are confident the student is expressing an incorrect idea.

But how indeed is an instructor to say ``yes, and…" in response to an untrue statement about the laws of physics? The key is that such statements invariably contain something of value that should be amplified. Students do not initiate interactions as empty vessels; they have fuzzy mental models~\cite{ten} and physical intuitions (``conceptual resources"~\cite{11} ) which form the basis for their questions. What they often lack is a sense of when to activate which resource, and which intuitions can be trusted. Consequently, when a student holds a view about the laws of physics that differs from the expert perspective, simply rejecting their belief or labeling it as ``wrong" can be less effective than an approach which harnesses and builds upon the correct aspects of their mental model~\cite{12}. The ``yes and" principle is a convenient reminder of this fact. 

Consider analyzing the forces on a ball which has been thrown through the air, after it has left the thrower’s hand (neglecting air resistance). Many students will produce a free body diagram like Figure 2, including an ongoing ``throw force." 

\begin{figure}
    \centering
    \includegraphics[width=0.4\textwidth]{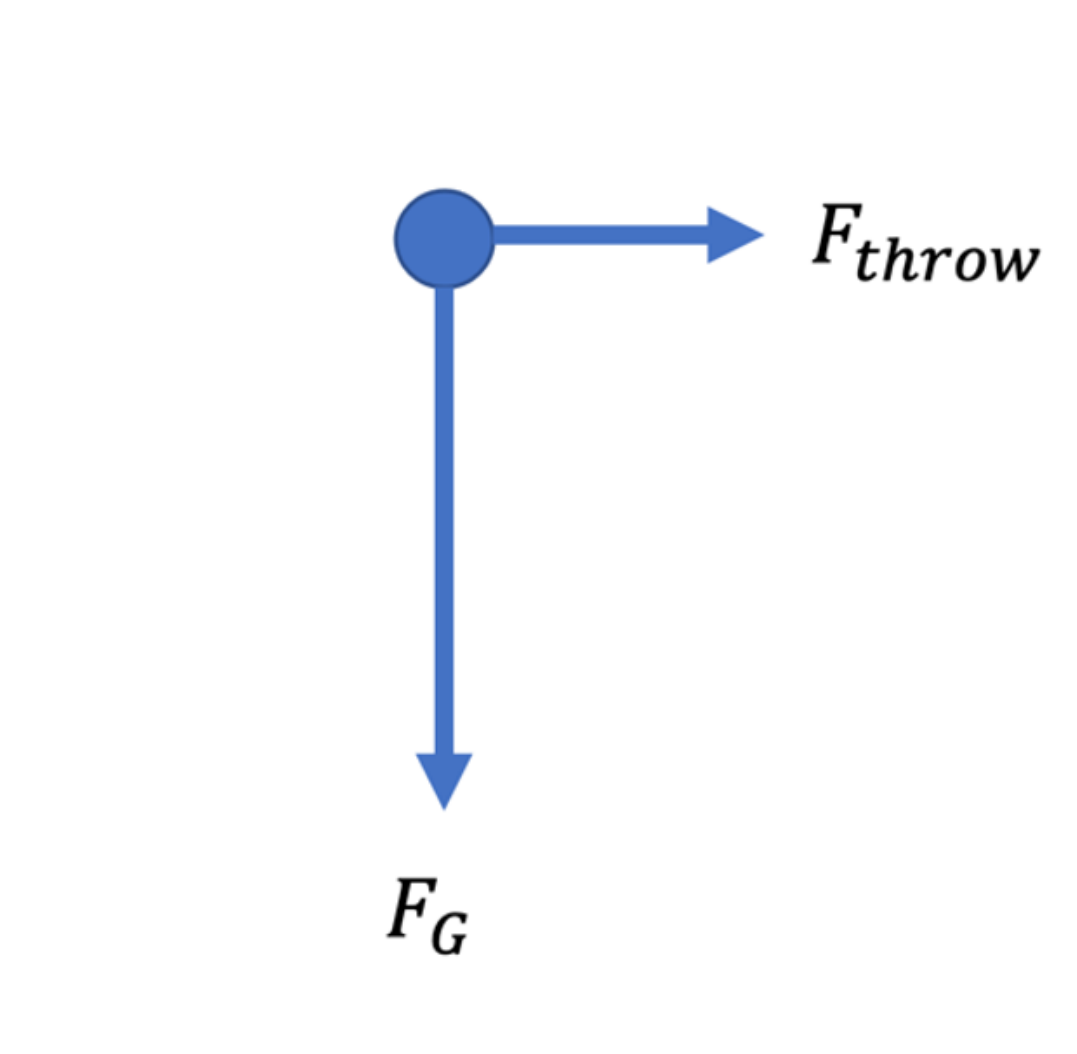}
    \caption{ A common but incorrect free body diagram for a ball which has been thrown through the air (neglecting air resistance). The presence of the nonexistent ``throw force" ($F_{throw}$) represents a misunderstanding of Newton’s laws, but is not devoid of pedagogical value.}
    \label{fig:throw}
\end{figure}

Suppose we ask a student why they drew the horizontal ``throw force," and they respond ``I knew there had to be something that kept the ball moving forward." It is tempting to respond, ``Ah but no, there’s no need for a force to explain an ongoing forward motion!" Of course, ``But no" is the antithesis of the improvisational rule suggested here. A much better response might be ``Ah yes! We do need to think about why the ball continues moving forward even after it has left my hand. And I think that actually, one of Newton’s laws might already explain this. Can anyone tell me which one?"

This second response provides the student a greater wealth of feedback: there is a concept they are correct about, but it connects differently than they thought to the rest of their knowledge. The student can update their web of conceptual resources accordingly, and our comments will help them find the right context in which to use them in the future~\cite{13}. Perhaps just as importantly, the ``yes, and" response also functions as an explicit recognition of the student’s knowledge, which over time can have important impacts on their sense of physics identity~\cite{14}. 

\section{Lesson 3: The ``Button"}

Since an improvised scene has not been planned, there is no guarantee that it will reach a satisfying and natural resolution. An important goal in many improv formats is finding a strong concluding line--called variously the ``button," the ``tag" or the ``out"~\cite{15}--which can give the scene a valuable sense of cohesion and identity. Before the ``button," the scene may just some business about a knight and a dragon. Afterwards, it might become the touching story of how a knight and a dragon became unexpected friends.

Similarly, a conversation with a physics student can easily trail off with a phrase like ``does that answer your question?"—but we should strive for more than this. As with an improv scene, a clear conclusion can allow the entire interaction to take on a more significant meaning. Students tend to remember unexpected results, but what they remember and whether it enhances comprehension depends heavily on instructional context~\cite{16}. Research also indicates that, while experts make efficient use of general patterns and broad categorizations when solving physics problems, novices are not so quick to make this leap~\cite{17, 18}. Accordingly, an expert instructor may feel that they have effectively made a general point, while the student may not recognize that it stands for a broader proposition. Carefully constructed activities making use of frameworks like the ``Elicit-Confront-Resolve" paradigm can help ensure that surprising results lead students to useful conclusions~\cite{19}. But in less structured interactions, this is harder to achieve. 

\begin{figure}
    \centering
    \includegraphics[width=0.45\textwidth]{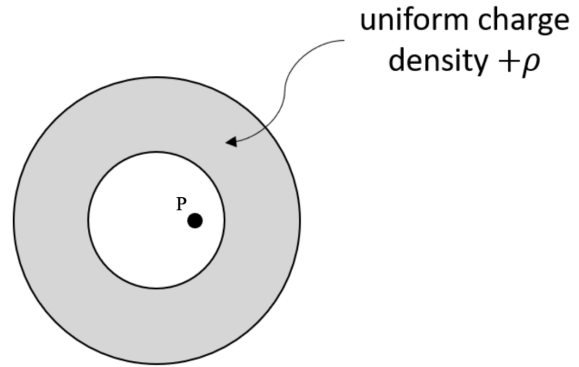}
    \caption{A notoriously counter-intuitive problem for students is to identify the magnitude of the electric field within a uniformly-charged insulating spherical shell, particularly at a point away from the center like point $P$.}
    \label{fig:shell}
\end{figure}

In such cases, we should strive to improvise a clear ``button" that emphasizes a useful takeaway message. Consider a famously counterintuitive problem regarding the electric field within a uniformly charged, spherically-symmetric insulating shell (Fig~\ref{fig:shell}). In situations of this type, many students find it very surprising that the electric field is zero everywhere within the shell~\cite{20}. 

With such an unintuitive result, a student is bound to remember discussing the problem with their instructor. But if the conversation ends with no particular button (``does that make sense?") the student may feel that the point was merely to find an answer in a very specific case. It is important to reach for a deeper level of generality—and yet we must be careful. ``Now you can see how easily your physical intuition can be led astray!" may be true, but it may not be desirable to leave students with the impression that their intuitions are always wrong. Perhaps better: ``Look how powerful Gauss’s law is in cases where intuition and superposition are not enough. It’s always worth trying Gauss’s law in situations with significant symmetry." This button would allows the memorable result to serve as a broad reminder for students who may overlook Gauss’s law as a problem-solving tool. 

Of course, a very different message may be appropriate depending on the student and the context. The important thing is that the takeaway message be chosen intentionally, even though it was not planned in advance. Improv teaches, ``to help find an end to the scene, look to the beginning"~\cite{ten}, since the ``meaning" of a scene is frequently contained in what has changed or evolved for the characters. In the same manner, a satisfying button for a student interaction often reaches back to the student’s initial premise to explicitly highlight the ways their mental models have been updated. 

\section{Conclusions}
The principles of improv theater teach many similar lessons to the literature of physics education research, but hold particular value by showing how these lessons should apply in the unplanned space of student-instructor interactions. In doing so they provide a structure and purpose to such interactions to keep them purposeful, beneficial, and memorable. Active listening allows the conversation to begin with both student and instructor headed in the same direction. The ``Yes-And" principle ensures that subsequent steps progress the conversation towards comprehension by building on shared knowledge. And finding a ``button" allows the whole interaction to underscore some more general principle.  

While I find that these three lessons form the backbone of my interactions with students, there are many other common improv principles that could have been written up here (such as ``show, don’t tell" and ``make accidents work"~\cite{15}). I encourage the interested reader to find more lessons that resonate with their teaching styles, through organizations that provide improv-based science communication training~\cite{one,21}, literature on improvisation in higher education~\cite{22,23}, or directly from the literature of improv theater itself~\cite{two, five, 15}.

\bibliographystyle{unsrt}
\bibliography{references}
\end{document}